\documentclass[conference]{IEEEtran}
\IEEEoverridecommandlockouts

\usepackage{hhline}
\usepackage{bm}
\usepackage{color}
\usepackage{longtable}	
\usepackage{multicol}
\usepackage{float}
\usepackage{multirow}

\usepackage{cite}
\usepackage{amsmath,amssymb,amsfonts}
\usepackage{algorithmic}

\usepackage{textcomp}
\usepackage{diagbox}
\usepackage{xcolor}
\usepackage{graphicx}
\usepackage{nicefrac}

\ifCLASSOPTIONcompsoc
\usepackage[caption=false,font=normalsize,labelfon
t=sf,textfont=sf]{subfig}
\else
\usepackage[caption=false,font=footnotesize]{subfi
g}
\fi
\pdfoutput=1

\def\BibTeX{{\rm B\kern-.05em{\sc i\kern-.025em b}\kern-.08em
    T\kern-.1667em\lower.7ex\hbox{E}\kern-.125emX}}

\newcommand{\E}{\mathbb{E}}

\newcommand{\Var}{\mathrm{Var}}

\newcommand\Tstrut{\rule{0pt}{2.4ex}}         
\newcommand\Bstrut{\rule[-0.9ex]{0pt}{0pt}}

\begin{document}

\IEEEoverridecommandlockouts
\IEEEpubid{\begin{minipage}[t]{\textwidth}\ \\[10pt]
\centering\normalsize{978-1-5386-8218-0/19/\$31.00 \copyright 2019 IEEE}
\end{minipage}}

\title{The Effect of the Uncertainty of Load and Renewable Generation on the Dynamic Voltage Stability Margin\\
\thanks{This work is supported by Natural Sciences and Engineering Research Council (NSERC) under Discovery Grant NSERC RGPIN-2016-04570 and by the Stavros S. Niarchos Foundation McGill Fellowship.
}
}

\author{\IEEEauthorblockN{Georgia Pierrou}
\IEEEauthorblockA{\textit{Department of Electrical and Computer Engineering} \\
\textit{McGill University}\\
Montreal, QC H3A 0E9, Canada \\
georgia.pierrou@mail.mcgill.ca}
\and
\IEEEauthorblockN{Xiaozhe Wang}
\IEEEauthorblockA{\textit{Department of Electrical and Computer Engineering} \\
\textit{McGill University}\\
Montreal, QC H3A 0E9, Canada\\
xiaozhe.wang2@mcgill.ca}
}


\maketitle

\begin{abstract}

In this paper, the impact of stochastic load and renewable generation uncertainty on the dynamic voltage stability margin is studied. Stochastic trajectories describing the uncertainty of load, wind and solar generation have been incorporated in the power system model as a set of Stochastic Differential-Algebraic Equations (SDAEs). A systematic study of Monte Carlo dynamic simulations on the IEEE 39-Bus system has been conducted to compute the stochastic load margin with all dynamic components active. Numerical results show that the uncertainty of both demand and generation may lead to a decrease on the size of the dynamic voltage stability margin, yet the variability of renewable generators may play a more significant role. Given that the integration of renewable energy will continue growing, it is of paramount importance to apply  stochastic and dynamic approaches in the voltage stability study. 
\end{abstract}
\vspace{-6pt}

\section{Introduction}

The growing load demand along with the increased penetration of intermittent renewable energy sources (RES) push power systems closer to their limits. The volatile nature of demand and generation is becoming a challenging issue for power system voltage stability. To ensure secure operation and power quality, uncertainty management requires more effort.

Research on static voltage stability under uncertainty has attracted significant attention. In \cite{Haesen09}, a probabilistic formulation for the voltage stability margin is proposed based on a Continuation Power Flow (CPF) algorithm that takes into account the stochastic power injections. In \cite{Ma11}, it is shown that the variability and the penetration level of wind generation may reduce the static loadability limit. A probabilistic analysis in \cite{Bu12} shows that the stochastic variation of wind generation can cause the system to lose stability even though the deterministic system is stable. The priority ranking of uncertain parameters in a network with RES using PV curves has been done in \cite{Qi17}. Polynomial Chaos Expansion and Low-Rank Approximation have been used in \cite{Sheng18}, \cite{Sheng19} to assess the available delivery capability for power systems with renewable variations.

However, limited work has been done to incorporate the randomness into \textit{dynamic} voltage stability. Compared to static approach, dynamic methods are shown to be more accurate to calculate the voltage stability margin since load dynamics and control actions can be incorporated \cite{Kundur93}. Stochastic dynamic power system  models have been proposed in \cite{Milano13} --\cite{Wang17} for voltage  stability study. It has been shown in \cite{Wang17} that there are critical cases in which the conventional deterministic model may fail to provide accurate results in the voltage stability assessment. The authors of \cite{Qiu08} used bifurcation theory to analyze how a stochastic load model may affect the voltage profile. Nonetheless, the conclusions are based on a single simulation of the stochastic power system model. In \cite{Pierrou19}, numerical results have revealed the impacts of stochastic load fluctuations on the dynamic load margin. Yet, the uncertainty of RES was not considered. In sum, a systematic study of the impact of both load and RES uncertainty on the dynamic load margin by Monte Carlo time domain simulations is lacking. 

In this paper, we attempt to systematically study the effect of stochastic load, wind and solar generation variations on dynamic voltage stability. To this end, uncertainty is incorporated in the power system model as a set of Stochastic Differential-Algebraic Equations (SDAEs). Time domain simulations are implemented to calculate the voltage stability margin. Numerical study of Monte Carlo simulations on the IEEE 39-bus system shows that the voltage in the stochastic model may collapse earlier, leading to a smaller dynamic load margin. The results indicate that the stochastic nature of both demand and generation may affect dynamic voltage stability, whereas the variability of renewable generation may play a more important role. 
As a result, to ensure power system secure operation, it is crucial that the uncertainty is carefully considered and stochastic dynamic approaches are adopted for voltage stability analysis. Particularly, this work may represent the first attempt to reveal the impacts of stochasticity from loads and renewable energy to the dynamic voltage stability margin with the implementation of time domain integration.


\section{Stochastic Dynamic Power System Model}

To incorporate the uncertainty in power system dynamics, power system is modeled as a set of SDAEs as follows:
\begin{equation}
\label{eq:sdae_full}
\begin{gathered}
\bm{\dot{x}} = \bm{f}(\bm{x},\bm{y},\bm{p}, \bm{\eta}) \\
\bm{0} = \bm{g}(\bm{x},\bm{y},\bm{p})
\end{gathered}
\end{equation}
where $\bm{x}$ are the state variables, e.g. rotor speeds and rotor angles of synchronous machines, the dynamic states of loads, etc.; $\bm{y}$ are the algebraic variables, e.g., bus voltage magnitudes and angles; $\bm{p}$ are the system parameters, e.g., load power at buses; $\bm{f}$ are the differential equations; $\bm{g}$ are the algebraic equations; $\bm{\eta}$ is the vector of stochastic perturbations, which can be modeled as a vector Ornstein-Uhlenbeck process \cite{Milano13}:

\begin{equation}
\label{eq:ouprocess}
\dot{\bm{\eta}} = -A_{\bm{\eta}}{\bm{\eta}} + \sigma B_{\bm{\eta}}{\bm{\xi}}, \quad t \in [0,T]
\end{equation}
where $A_{\bm{\eta}}=\mbox{diag}([\alpha_{1},...,\alpha_{k}])$ and is positive definite; $\sigma$ is the noise intensity; $B_{\bm{\eta}}=\mbox{diag}([\beta_{1},...,\beta_{k}])$ denotes the relative strength between perturbations; $\int_{0}^{t} \bm{\xi}(u) du$ is a $k-$dimensional Brownian motion.

Choosing the initial condition $\eta_i(0) \sim \mathcal{N}(0,(\sigma\beta_i)^2/2\alpha_i)$, for each stochastic process $\eta_i$ we have:
\begin{itemize}

\item $\E[\eta_i(t)]=0, \quad  \forall t \in [0,T],$
\item $\Var[\eta_i(t)]=(\sigma\beta_i)^2/2\alpha_i, \quad \forall t \in [0,T],$

\item $\mbox{Aut}[\eta_i(t_{m}),\eta_{i}(t_{n})] = e^{-\alpha_{i} |t_{n}-t_{m}|}, \quad \forall t_{m},t_{n} \in [0,T]$.

\end{itemize}
According to the Implicit Function Theorem, if the algebraic Jacobian matrix $g_{\bm{y}}$ is non-singular, $\bm{y}$ can be eliminated as follows \cite{Cutsem}:
\begin{eqnarray}
\label{eq:sde_model_withouty_1}
\dot{\bm{x}} &=& \bm{H}(\bm{x},\bm{p}, \bm{\eta}) \\
\dot{\bm{\eta}} &=& -A_{\bm{\eta}}{\bm{\eta}} +\sigma B_{\bm{\eta}}{\bm{\xi}}
\label{eq:sde_model_withouty_2}
\end{eqnarray}
Equations (\ref{eq:sde_model_withouty_1})-(\ref{eq:sde_model_withouty_2}) can be written in the compact form:
\begin{equation}
\label{eq:sde_model_u}
\dot{\bm{u}} = \bm{G}(\bm{u},\bm{p})+\sigma B \bm{\xi}
\end{equation}
where $\bm{u}=\begin{bmatrix} {\bm{x}}, {\bm{\eta}}
\end{bmatrix}^T$ and $B=\begin{bmatrix} \bm{0}, B_{\bm{\eta}} \end{bmatrix}^T$.

The detailed representation for the load and RES uncertainty in the form of SDAEs can be found subsequently.

\subsection{The Stochastic Dynamic Load Model}

Load dynamic response is one of the key mechanisms of power system voltage stability as it drives the dynamic evolution of voltages. In this paper, load dynamics are represented by an aggregated self-restoring load, also known as the Exponential Recovery Load (ERL) model, which has been proposed to naturally represent the most common types of loads depending on the selected time constant \cite{Nguyen16}.  The active and reactive consumption of the ERL are as follows:
\begin{equation}
\label{eq:erl_power}
\begin{gathered}
p=x_p/T_p+p_t\\
q=x_q/T_q+q_t
\end{gathered}
\end{equation}
where $x_p$ and $x_q$ are the state variables given by:
\begin{equation}
\label{eq:erload}
\begin{gathered}
\dot{x}_p=-x_p/T_p+p_s-p_t \\
\dot{x}_q=-x_q/T_q+q_s-q_t
\end{gathered}
\end{equation}
${T_{p}}$ and ${T_{q}}$ are the corresponding power time constants; ${{p}_{s}}$ and ${{p}_{t}}$ are the static and transient active power absorptions; ${{q}_{s}}$ and ${{q}_{t}}$ are the static and transient reactive power absorptions.

Since $p_s$, $p_t$, $q_s$ and $q_t$ are voltage dependent, Ornstein-Uhlenbeck (\ref{eq:ouprocess}) stochastic load variations can be incorporated into the aggregated load model as follows \cite{Milano13}, \cite{Pierrou19}, \cite{Wangxz:2017}, \cite{Tang19}:
\begin{equation}
\label{eq:exp_rec_load_sde}
\begin{gathered}
{{p}_{s}}={({p}_{0}+\eta_{i}(t))}(\frac{V}{{V}_{0}})^{{\alpha}_{s}} \quad {{p}_{t}}={({p}_{0}+\eta_{i}(t))}(\frac{V}{{V}_{0}})^{{\alpha}_{t}} \\
{{q}_{s}}={({q}_{0}+\eta_{i}(t))}(\frac{V}{{V}_{0}})^{{\beta}_{s}} \quad
{{q}_{t}}={({q}_{0}+\eta_{i}(t))}(\frac{V}{{V}_{0}})^{{\beta}_{t}} \\
\end{gathered}
\end{equation}
where ${{p}_{0}}$ and ${{q}_{0}}$ are the nominal active and reactive load power; $\eta_i$ describes the stochastic perturbations around the nominal power; ${{\alpha}_{s}}$, ${{\beta}_{s}}$, ${{\alpha}_{t}}$ and ${{\beta}_{t}}$ are exponents related to the steady state and the transient load response, respectively; ${{V}_{0}}$ is the nominal bus voltage.






\subsection{The Stochastic Wind Speed Model}
\label{stochwindspeed}
A continuous wind speed model based on Stochastic Differential Equations (SDEs) has been developed in \cite{Wang17}, \cite{Minano13} to generate wind speed trajectories of any time scale that carry the statistical properties of real wind speed data. Briefly speaking, given the two-parameter Weibull distribution and the autocorrelation function derived from wind speed measurements, an autocorrelated Weibull distributed  wind speed model can be developed by applying a memoryless transformation to the Ornstein-Uhlenbeck process $\eta_{w}$ \eqref{eq:ouprocess}:
\begin{equation}
\label{eq:weibull_ou_proc}
w(t) = g(\eta_{w}(t)) = F_w^{-1}(\Phi(\frac{\eta_{w}(t)}{\beta_{w}/\sqrt{2\alpha_{w}}}))
\end{equation}
$F_w$ is the Weibull cumulative distribution function:
\begin{equation}
\label{eq:weibull_cdf}
F_w(z) = 1 - e^{-(\nicefrac{z}{\lambda})^k} \quad \forall z>0
\end{equation}
where ${\lambda >0}$ is the scale parameter and ${k>0}$ is the shape parameter of the Weibull distribution. ${\Phi}$ is the Gaussian cumulative distribution function:
\begin{equation}
\label{eq:gaussian_cdf}
\Phi(\frac{z-E[z]}{\sqrt{Var[z]}}) = \frac{1}{2}(1 + \mbox{erf}(\frac{z-E[z]}{\sqrt{2Var[z]}})) \quad \forall z \in \mathbb{R}
\end{equation}
The wind speed processs $w(t)$ has the following statistical properties, where $\Gamma$ denotes the Gamma function:
\begin{itemize}
\item $\mu_{w}=E[w(t)]=\lambda\Gamma(1+\frac{1}{k}), \quad \forall t \in [0,T]$
\item $\sigma_{w}^2=Var[w(t)]=\lambda^2\Gamma(1+\frac{2}{k})-\mu_{w}^2, \quad \forall t \in [0,T]$
\item $Aut[w(t_{m}),w(t_{n})]\approx e^{-\alpha_{w} |t_{n}-t_{m}|}, \quad \forall t_{m},t_{n} \in [0,T]$.
\end{itemize}

It is worth noting that the only parameters required for the model are $\lambda, k$ and $\alpha_{w}$ which are given from historical wind speed data, whereas $\beta_{w}$ does not affect the statistical properties above.

\subsection{The Stochastic Solar Irradiance Model}
\label{stochsolarirradiance}
Beta distribution has been typically proved to be a good fit for real solar irradiance measurements \cite{Ettoumi02}, \cite{Trashchenkov18}. Similarly to the Weibull distributed process for the wind speed, a Beta autocorrelated model for the solar radiation based on SDEs can be developed. In this case, we apply a memoryless transformation to the Ornstein-Uhlenbeck process $\eta_s$ \eqref{eq:ouprocess} to get a Beta distributed process for the solar irradiance, as follows:
\begin{equation}
\label{eq:beta_ou_proc}
s(t) = h(\eta_{s}(t)) = F_B^{-1}( \Phi(\frac{\eta_{s}(t)}{\beta_{s}/\sqrt{2\alpha_{s}}}))
\end{equation}
where  ${\Phi}$ is described in (\ref{eq:gaussian_cdf}) and ${F_B}$ is the Beta cumulative distribution function:
\begin{equation}
\label{eq:beta_cdf}
F_B(z,p,q) = \frac{\int_{0}^{z} t^{p-1}(1-t)^{q -1}dt}{B(p,q)} \quad 0 \leq z \leq 1 \quad p,q >0
\end{equation}
and the function ${B}$ is defined as follows:
\begin{equation}
B(p,q) = \int_{0}^{1} t^{p-1}(1-t)^{q -1}dt
\end{equation}
The resulting process $s(t)$ for the solar irradiance is an autocorrelated Beta distributed process with the following statistical properties:
\begin{itemize}
\item $\mu_{s}=E[s(t)]=\frac{p}{p+q}$
\item $\sigma_{s}^2=Var[s(t)]=\frac{pq}{(p+q)^2(p+q+1)}$
\end{itemize}
where $p, q$ are the shape parameters of the Beta distribution function that fits the solar irradiance measurements.

The relation among the solar irradiance and the active PV power is the following \cite{Sheng18}:
%
\begin{equation}
\label{eq:solar_p_val}
{{P}_{pv}}(t_{i})={{P}_{pv}}(s(t_{i}))=\left\{ \begin{array}{*{35}{l}}
\displaystyle \frac{{{s(t_{i})}^{2}}}{{{r}_{c}}{{r}_{std}}}{{P}_{r}} & 0\le s(t_{i})<{{r}_{c}}  \\
\displaystyle \frac{s(t_{i})}{{{r}_{std}}}{{P}_{r}} & {{r}_{c}}\le s(t_{i}) < {{r}_{std}}  \\
{{P}_{r}} & s(t_{i}) \ge {{r}_{std}}  \\
\end{array} \right.
\end{equation}
where ${{r}_{c}}$ is a specific radiation threshold level up to which a small increase in the radiation produces a significant increase in the PV output; ${{r}_{std}}$ is the solar radiation in the standard environment where further increase in radiation produces a relatively small change in the PV output; ${{P}_{r}}$ is the rated capacity of the solar installation.

\section{Voltage Stability Margin}

If there is no stochasticity, i.e. $\sigma=0$ in (\ref{eq:sde_model_u}), the deterministic power system model is defined as:
\begin{equation}
\label{eq:ode}
\bm{\dot{u}}  = \bm{G}(\bm{u},\bm{p}(t))
\end{equation}
where $\bm{p}=\bm{p}(t)$ are slowly changing parameters, e.g. the active power $P$ and the reactive power $Q$ of loads which may gradually increase with respect to time. In power system voltage stability study, Voltage Dependent Load (VDL) models are commonly used \cite{Kundur93}, \cite{Cutsem}, \cite{PESreport}. We assume load power increases as follows:

\begin{equation}
\label{slowlychanging}
\begin{gathered}
    P=P_{0}(1+z(t))(\frac{V}{{V}_{0}})^{\alpha_P}\\
    Q=Q_{0}(1+z(t))(\frac{V}{{V}_{0}})^{\alpha_Q}
\end{gathered}
\end{equation}
where $z$ is a scaling factor increasing linearly with time.

As $z$ increases, a critical point may be reached and bifurcations may occur. At this critical stage, a smooth parameter change may result in sudden changes in the power system response, e.g. oscillatory behavior or change of stability and disappearance of equilibrium points. Voltage collapse has been revealed to be closely related to the Saddle-Node Bifurcation (SNB) \cite{Chiang89}, where two equilibrium points--one stable and one unstable--coalesce and disappear.
The SNB point corresponds to the nose of the PV curve. A crucial quantity  typically used as a security measure is the voltage stability margin, which can be quantified in terms of the MW distance between the current operating point and the SNB. In (\ref{slowlychanging}), the voltage stability margin can be defined as $z^\star P_{0}$, where $z^\star$ is the value of the scaling factor when SNB happens.

Unlike classical PV curves that can be directly calculated from CPF computations, in dynamic voltage stability study the calculation of the margin is based on time domain simulations. Specifically, a slow ramp increase in load power is simulated in time as in (\ref{slowlychanging}), representing what could be observed in a real system. 
The power system response is computed with all dynamic components active including control actions. The voltage stability margin is calculated as soon as it is ascertained that maximum load power has been reached. A useful indicator is the critical eigenvalue, i.e., the one with the smallest absolute value, that goes through a sign change at the point of instability \cite{Cutsem}.

\section{Numerical Results}
In this work, we plan to investigate the impact of load and generation uncertainty on the dynamic voltage stability margin. Different case studies are presented, where the statistic properties of the dynamic load margin have been calculated by 1000 Monte Carlo time domain simulations. The first study focuses on computing the dynamic load margin when stochastic dynamic load fluctuations are incorporated in the model. In the second study, renewable generation variations are added on top of the existing stochastic dynamic load fluctuations to show how the volatile nature of RES may further affect the dynamic voltage stability.

\begin{figure}[!tb]
\vspace{-10pt}
    \centering
    \includegraphics[width=3.2in ,keepaspectratio=true,angle=0]{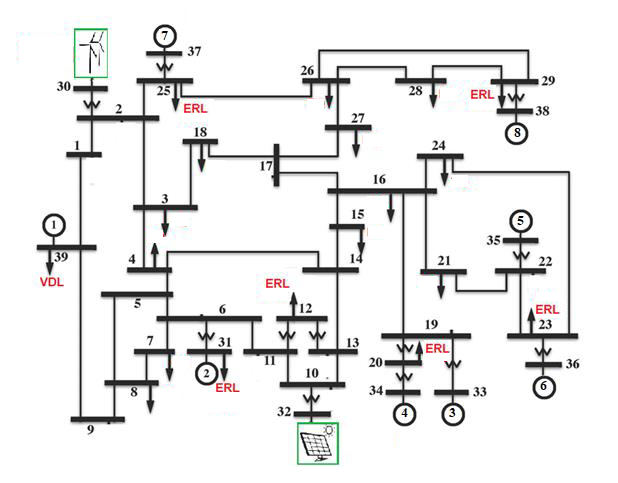}
    \caption{The IEEE 39-Bus 10-Generator New England system.}
    \label{ieee39}
\end{figure}

The IEEE 39-Bus 10-Generator New England system has been used to conduct the numerical study. The topology of the system can be found in Fig. \ref{ieee39}.
As shown in Fig. \ref{ieee39}, conventional synchronous generators at Buses 30 and 32 have been replaced by a Doubly-Fed Induction Generator (DFIG) driven by a wind source and a Solar Constant PQ Generator (SPQ) respectively, representing a RES penetration level of approximately $15\% $. In addition, there are 8 synchronous generators that are equipped with Turbine Governors (TGs), Automatic Voltage Regulators (AVRs) and Over-Excitation Limiters (OXLs). Continuous Load Tap Changers (LTCs) are installed at Buses 20 and 31 with initial tap delays 20s and subsequent tap delays 5s. ERLs replace the conventional PQ loads at the participating buses (Buses 12, 20, 23, 25, 29, 31). The load at Bus 39 is considered to be the slowly changing parameter, modeled as a VDL that increases 0.005 of its nominal power per second, as in (\ref{slowlychanging}).

All the simulations were implemented using PSAT Toolbox \cite{Milano05}.  Ornstein Uhlenbeck processes were generated with the Euler-Maruyama method. The integration time step was $\Delta t=0.05$s.

\subsection{Impact of Stochastic Load Fluctuations}
\label{numstudy1}
In the first study, we focus on the effect of stochastic dynamic load fluctuations on the dynamic voltage stability margin. To this end, stochastic load fluctuations with $\sigma=0.05, \alpha_{i}=1, \beta_{i}=\sqrt{2\alpha_{i}}$ are applied to the dynamic ERLs as in (\ref{eq:exp_rec_load_sde}), whereas the amount of renewable power generated by the DFIG and the SPQ is regarded as constant.

\begin{figure}[!htb]
 \centering
 \includegraphics[width=2.6in ,keepaspectratio=true,angle=0]{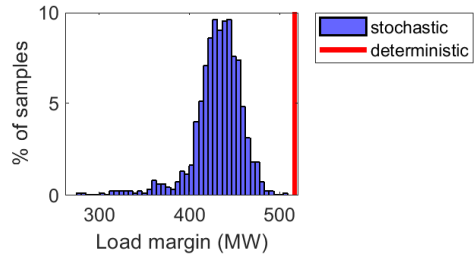}
 \caption{The distribution of the dynamic voltage stability margin when stochastic load fluctuations are considered.}\label{loadfluctuations_histogram}
 \vspace{-5pt}
 \end{figure}

For the deterministic system, i.e., $\sigma$ = 0 in (\ref{eq:sde_model_u}), the load margin is 516.45 MW. The distribution of the load margin for the stochastic system is depicted in Fig. \ref{loadfluctuations_histogram}. Also, the statistics of the margin  (mean value $\mathbb{E}(z^\star P_0)$, standard deviation $\mathrm{Std}(z^\star P_0)$  and the 90\% interval) are reported in Table \ref{loadfluctuations_table}. From the results we can observe that the stochastic dynamic load fluctuations affect dynamic voltage stability, leading to a decrease in the dynamic stability margin for all the samples of the stochastic power system model. More specifically, the mean value for the stochastic margin is $E(z^\star P_{0})=431.78$ MW, indicating a 16.4\% decrease comparing to the deterministic case.

The results of the study clearly show that the uncertainty brought about by stochastic load fluctuations may affect the dynamic load margin when renewable generation is regarded as constant. Therefore, an important question is what may happen if the stochasticity from the RES is additionally considered in the dynamic voltage stability study.


\begin{table}[!hb]
\centering
  \caption{The statistics and the 90\% interval of the dynamic load margin variation considering stochastic load fluctuations}\label{loadfluctuations_table}
  \setlength{\tabcolsep}{2pt}
  \begin{tabular}{|c |c |c| }
\hhline{|=|=|=|}
\hline
 $\mathbb{E}(z^\star P_0)$ (MW) & $\mathrm{Std}(z^\star P_0)$&  $d_{90 \%}$ (MW) \Tstrut\Bstrut\\
  \hline
431.78&26.54&406.02 \\
\hhline{|=|=|=|}
  \end{tabular}

 \end{table}

\subsection{Impact of Stochastic Load and Renewable Generation Variations}
In the second study, we focus on the impact of stochastic renewable generation variations. In addition to the stochastic load fluctuations applied as in Section \ref{numstudy1}, the volatile nature of the wind speed and the solar irradiance is considered. The stochastic modeling for the wind speed is implemented as described in Section \ref{stochwindspeed} and the parameters derived from real measurements \cite{Minano13} are $c=3.36, k=1.51, \alpha_w=0.2575/3600 (\nicefrac{1}{s}), \beta_{w}=\sqrt{2\alpha_{w}}$. Stochastic solar irradiance trajectories are generated according to Section \ref{stochsolarirradiance} and the associated parameters are $p=1.11, q=0.73$ \cite{Ettoumi02}, $\alpha_s=0.2231/3600 (\nicefrac{1}{s})$ \cite{Gordon88}, $\beta_{s}=\sqrt{2\alpha_{s}}$, $r_c=150 W/m^2$, $r_{std}=1000 W/m^2$ \cite{Sheng18}.

\begin{figure}[!ht]
 \vspace{-10pt}
 \centering
 \includegraphics[width=2.6in ,keepaspectratio=true,angle=0]{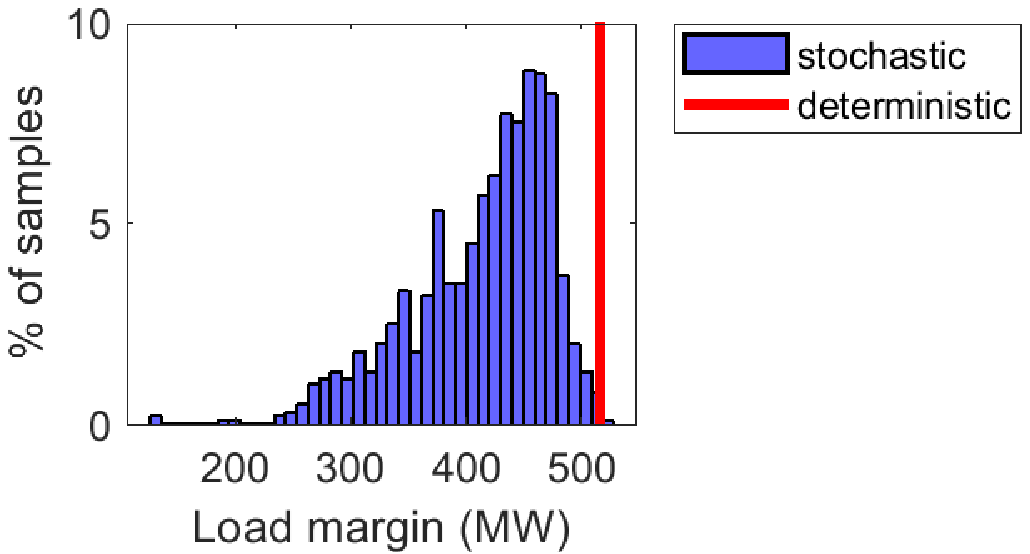}
 \caption{The distribution of the dynamic voltage stability margin when stochastic load and renewable generation fluctuations are considered.}\label{renewablegenerations_histogram}
 \vspace{-10pt}
 \end{figure}

The distribution of the load margin for the stochastic system with load and renewable variations is illustrated in Fig. \ref{renewablegenerations_histogram}. Also, the statistics of the margin (mean value $\mathbb{E}(z^\star P_0)$, standard deviation $\mathrm{Std}(z^\star P_0)$  and the 90\% interval) can be found in Table \ref{renewablegenerations_table}. It can be observed that the stochastic renewable generation seems to further affect the size of the margin comparing to the results of Section \ref{numstudy1}. The mean value of the samples is 413.52 MW, corresponding to a reduction of almost $20\%$ from the deterministic margin under approximately 15\% penetration of RES. Compared to the previous case study, it can be seen that the variability of RES may play a more important role than that of loads. Given that the trend of increasing integration of RES will continue in the future, it is of paramount importance to incorporate the variability of RES in the voltage stability assessment.

\begin{figure}[!ht]
\vspace{-5pt}
 \centering
 \includegraphics[width=2.6in ,keepaspectratio=true,angle=0]{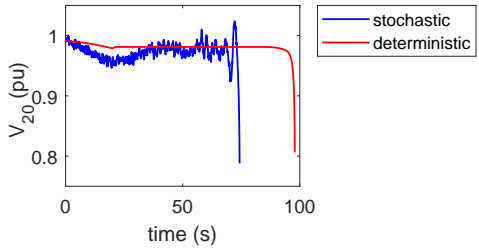}
 \caption{Comparison of one stochastic and the deterministic trajectory of the voltage magnitude at Bus 20.}\label{voltage20_renewable}
 \vspace{-10pt}
 \end{figure}

Furthermore, Fig. \ref{voltage20_renewable} presents the voltage magnitude $|V_{20}|$ at Bus 20 for the deterministic case and for one stochastic realization. It can be observed that the voltage collapse occurs earlier for the stochastic system, resulting in a smaller size of the voltage stability margin. It is worth noting that before the collapse the voltage at Bus 20 shows no gradual decrease due to continuous LTC operation, indicating that conventional sensitivity approach of detecting voltage collapse without time domain simulation  may fail. 

 The above results show that the forecasted evolution of the wind speed and solar irradiance play important roles when computing the dynamic load margin. Hence, the importance of carrying out dynamic and stochastic approaches is reinforced, since they seem to provide a more accurate representation for the time evolution of power system variables. Such information can be found useful towards the determination of secure operation limits in power systems with high penetration of RES experiencing some load stress.

\begin{table}[!htb]
\vspace{-8pt}
\centering
  \caption{The statistics and the 90\% interval of the dynamic load margin variation considering stochastic load and renewables}\label{renewablegenerations_table}
  \setlength{\tabcolsep}{2pt}
  \begin{tabular}{|c |c |c| }
\hhline{|=|=|=|}
\hline
 $\mathbb{E}(z^\star P_0)$ (MW) & $\mathrm{Std}(z^\star P_0)$&  $d_{90 \%}$ (MW)  \Tstrut\Bstrut\\
  \hline
413.52&60.14&326.03 \\
\hhline{|=|=|=|}
  \end{tabular}
 \end{table}

\section{Conclusions}
In this paper, we have studied the impact of stochastic load and renewable generation uncertainty on the dynamic voltage stability margin. Stochastic power system modeling has been developed based on SDAEs. An exhaustive study on the IEEE 39-bus system including Monte Carlo time domain simulations has been conducted to compute the stochastic load margin with all dynamic components active. Numerical results indicate that the volatile nature of both demand and renewable generation may lead to a decrease in the size of the dynamic load margin. Therefore, the stochastic characteristics should be carefully incorporated in the voltage stability study and dynamic methods should be applied to investigate their impact on the time evolution of the power system. Particularly, this work may represent the first attempt to reveal the effect of uncertainty on the dynamic margin using dynamic simulations. Future research may focus on identifying the most critical uncertain parameters affecting dynamic voltage stability.

\section*{Acknowledgment}
Author G.P. would like to thank Dr. Hao Sheng for providing the algorithm to test the stochastic generation trajectories.

\vspace{-15pt}
\bibliography{isgt2019}
\vspace{12pt}

\end{document}